# Self-Modulation Doping Effect in the High-Mobility Layered Semiconductor $Bi_2O_2Se$

Huixia Fu[1][*], Jinxiong Wu[2][*], Hailin Peng[2], and Binghai Yan[1]

1. Department of Condensed Matter Physics, Weizmann Institute of Science, Rehovot, 7610001, Israel
2. Center for Nanochemistry, Beijing Sciences and Engineering Centre for Nanocarbons, College of Chemistry and Molecular Engineering, Peking University, Beijing 100871,China

Email: binghai.yan@weizmann.ac.il

* These two authors contributed equally to the work.

**ABSTRACT:**
**Recently an air-stable layered semiconductor $Bi_2O_2Se$ was discovered to exhibit an ultrahigh mobility in transistors fabricated with its thin layers. In this work, we explored the mechanism that induces the high mobility and distinguishes $Bi_2O_2Se$ from other semiconductors. We found that the electron donor states lie above the lowest conduction band. Thus, electrons get spontaneously ionized from donor sites (e.g., Se vacancies) without involving the thermal activation, different from the donor ionization in conventional semiconductors. Consequently, the resistance decreases as reducing the temperature as observed in our measurement, which is similar to a metal but contrasts to a usual semiconductor. Furthermore, the electron conduction channels locate spatially away from ionized donor defects (Se vacancies) in different van der Waals layers. Such a spatial separation can strongly suppress the scattering caused by donor sites and subsequently increase the electron mobility, especially at the low temperature. We call this high-mobility mechanism self-modulation doping, i.e. the modulation doping spontaneously happening in a single-phase material without requiring a heterojunction. Our work paves a way to design novel high-mobility semiconductors with layered materials.**

High mobility devices and materials are significant for both the fundamental research and semiconductor technology. A well-known method to realize the high mobility is the modulation doping based on a heterostructure that spatially separates carrier from ionized impurities [1]. This mechanism was also theoretical proposed to design highly conductive semiconductor nanowires recently [2,3]. In recent years, two-dimensional (2D) materials including graphene [4-9] and three-dimensional topological semimetals [10,11] provides new high-mobility platforms with ultrafast Dirac electrons. Very recently, an emerging layered semiconductor, bismuth oxychalcogenide ($Bi_2O_2Se$), was found to exhibit ultrahigh electron mobility in fabricated devices [12]. Soon this material was fabricated into ultrafast, highly-sensitive infrared photodetectors [13] and magnetoresistance devices [14], and was predicted to be a candidate for the ferroelectric property [15].

The $Bi_2O_2Se$ compound attracts great research attention for designing novel devices, while the fundamental origin of its high mobility remains unexplored. Similar layered chalcogenides, such as the well-known topological insulator [16,17] $Bi_2Se_3$, usually suffer from the self-doping (e.g., Se vacancies) and exhibits very low mobility [18-20]. Furthermore, the resistance of a usual semiconductor decreases as increasing the temperature [21], because free carriers require the thermal energy to get ionized from donor/acceptor sites. In contrast, the resistance of $Bi_2O_2Se$ exhibits an opposite trend to the temperature [12,22]. The uniqueness of this material motivates us to investigate its electronic structure to understand its transport properties.

In ordinary semiconductors, it is known as the defect states are normally located in the band gap and lower than the empty conduction bands as shown in Fig. 1a. In order to achieve the conducting channels, the excess electrons should be excited to the next conduction band by temperature. Free electrons display the higher density at a higher temperature and thus, the resistance will decease sharply with increasing temperature. In contrast, $Bi_2O_2Se$ samples exhibit a different *R-T* dependence from ordinary semiconductors (see Fig. 1b).

In this article, we have performed experimental and theoretical studies on the electronic and transport properties of $Bi_2O_2Se$. We have found that Se vacancies ($V_{Se}$) and Se–Bi antisites ($Se_{Bi}$) are crucial donors for electron carriers in this material. Because the $V_{Se}$ distributes in the Se layer while the wave function of conduction electrons locates in the ($Bi_2O_2$) layer in the lattice, the scattering of $V_{Se}$ to electron carriers is much weak. However, $Se_{Bi}$ scatters electrons strongly, since it distributes in the same ($Bi_2O_2$) layer as the conduction band (Fig. 1c). Interestingly, as convinced by the formation energy calculations, the relative amount of $V_{Se}$ and $Se_{Bi}$ can be readily adjusted by changing the Se-richness during synthesis. $V_{Se}$ is the major defect in the Se-poor condition while $Se_{Bi}$ is the major one in the Se-rich condition, which well explained the low-temperature mobility variation of $Bi_2O_2Se$ synthesized under different Se-richness condition. Further, we found that the donor levels of both $V_{Se}$ and $Se_{Bi}$ lie above the conduction band bottom in energy by calculations. Such a band structure induces the automatic ionization of free electrons from these defects without overcoming an activation barrier (Fig. 1b). It explains the metal-like temperature-dependence of the resistance measured.

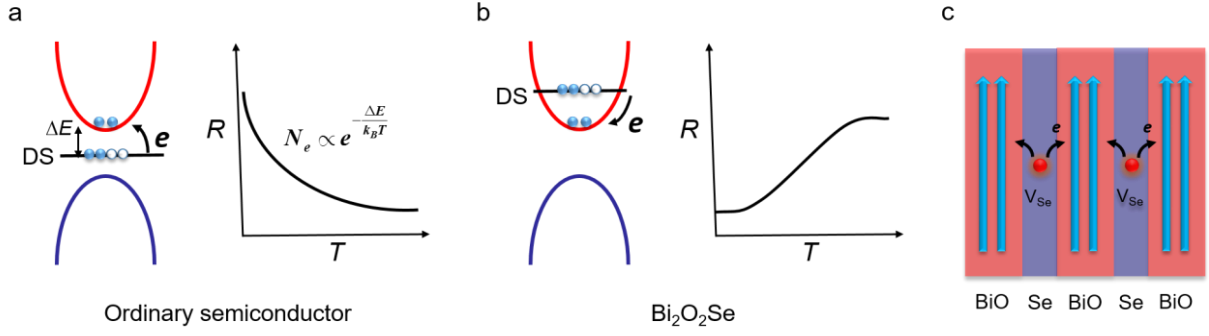

Figure 1. Energy level for defect state (DS), electron doping process and temperature-dependent resistance for ordinary semiconductors (a) and $Bi_2O_2Se$ (b). (c) Diagram of spatial separation between conducting channels and donor vacancies ($V_{Se}$) in $Bi_2O_2Se$.

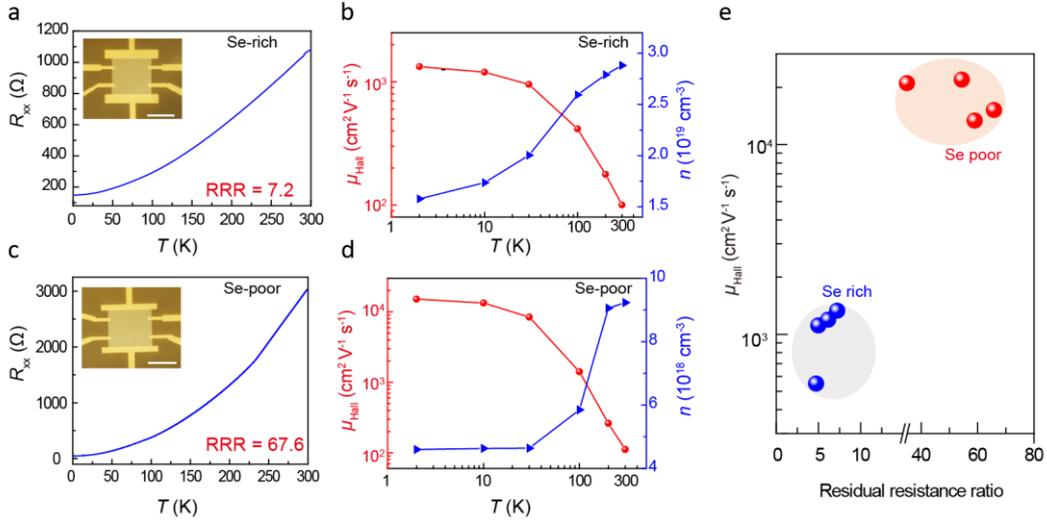

Figure 2. Electrical measurements of 2D $Bi_2O_2Se$ crystals synthesized under different Se-richness growth condition. (a, b) Typical temperature-dependence longitudinal resistance ($R_{xx}$) of 2D $Bi_2O_2Se$ crystals synthesized under relatively Se-rich (a) and Se-poor condition (b), showing significant difference on residual resistance ratios (RRR). The Se-richness is controlled by adjusting the ratio of $Bi_2Se_3$ and $Bi_2O_3$ that were used as co-evaporation sources (i.e. keep the amount of $Bi_2O_3$ constant, while changing the amount of $Bi_2Se_3$ independently). Inset: OM images of $Bi_2O_2Se$ Hall-bar devices fabricated on mica substrate. Both $Bi_2O_2Se$ devices have a similar thickness of 10 nm. Scale bar: 50 µm. (c, d) The corresponding temperature-dependence Hall mobility ($\mu_{Hall}$) and carrier density ($n$) of 2D $Bi_2O_2Se$ crystals synthesized under relatively Se-rich (c) and Se-poor condition (d). (e) Statistics for low-temperature Hall mobility (2 K) and residual resistance ratios of 2D $Bi_2O_2Se$ crystals synthesized

under different Se-richness growth condition. Much higher Hall mobility and residual resistance ratio were obtained on $Bi_2O_2Se$ crystals synthesized under a relatively Se-poor condition.

In the Se-richness controlled CVD growth experiments (for details, see experimental section in Supplementary Information), we found there are two types of 2D $Bi_2O_2Se$ samples with obviously distinct transport performances under different Se-richness growth conditions. The Se/Bi ratio is controlled by adjusting the ratio of $Bi_2Se_3$ and $Bi_2O_3$ that were used as co-evaporation sources (i.e. keep the amount of $Bi_2O_3$ as constant, while changing the amount of $Bi_2Se_3$ independently). As shown in Fig. 2a and 2b, both the Se-poor and Se-rich samples show a metal-like *R-T* behavior, namely longitudinal resistance ($R_{xx}$) decreases monotonously upon cooling down. This *R-T* feature is different from the ordinary semiconductor, whose resistance usually increase sharply upon cooling down (see Fig. 1a for example). Remarkably, the $Bi_2O_2Se$ samples synthesized under different Se-richness shows significant difference on residual resistance ratio (RRR, defined as defined as $R_{xx,\ 300\ K}/R_{xx,\ 2\ K}$), which is a key parameter to reflect the intrinsic quality of samples obtained. As shown in Fig. 2a and 2b, the typical RRR of Se-rich $Bi_2O_2Se$ (67.6) is about an order of magnitude higher the Se-poor one (7.2). A higher RRR usually indicates a higher Hall mobility. Fig. 2c and 2d showed the evolution of Hall mobility and carrier density as a function of temperature. The Hall mobility ($\mu_{Hall}$) is obtained from $\mu_{Hall} = (L/W)(G/ne)$, where *e* is the charge of an electron and *n* is the 2D charge density determined from Hall coefficient $R_H$ measurements ( $n = 1/eR_H$). Remarkably, the Hall mobility of both Se-rich and Se-poor samples increased monotonously as the temperature cools downs to 2 K, and this feature fits well with a phonon dominated charge transport mechanism. Interestingly, the Se-poor sample holds a significantly higher Hall mobility (>15, 000 $cm^2\ V^{-1}\ s^{-1}$) than the Se-poor sample (~1000 $cm^2\ V^{-1}\ s^{-1}$) at 2 K, while showing a similar room-temperature Hall mobility. It can be well explained by taking into consideration that the charge impurities (defects) scattering usually dominate the scattering events at low temperature, while phonon scattering dominates at room temperature [23-25]. In other words, our experimental results suggest, to some extent, the existence of some kinds of defects that may greatly depress the mobility of Se-rich $Bi_2O_2Se$. To further confirm this feature, we performed the statistics for low-temperature Hall mobility (2 K) and residual resistance ratios of 2D $Bi_2O_2Se$ crystals synthesized under different Se-richness growth condition (Fig. 2e). Obviously, the Se-poor samples indeed show much higher mobility and residual resistance ratio than the Se-rich one.

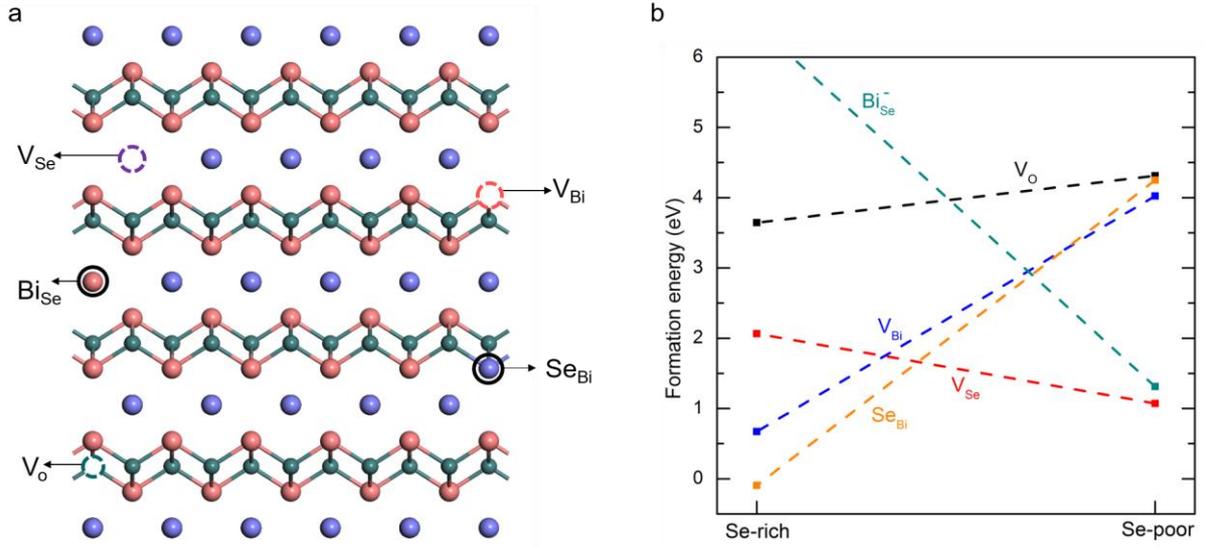

Figure 3. (a) Atomic structure of $Bi_2O_2Se$ with five possible defects. (b) The formation energy for these defects with respect to the chemical potential at different Se richness conditions.

To understand the origin of unusual resistance behavior and the high mobility, we have performed the fist-principles calculations [26] to explore the structural and electronic properties of $Bi_2O_2Se$ in the following. We first estimate the formation of possible defects as potential donors and acceptors with respect to the experimental condition. The free carriers in $Bi_2O_2Se$ are expected to come from lattice defects, since there is no specific impurities introduced in the experiment. As the atomic configuration displayed in Fig. 3a, $Bi_2O_2Se$ exhibits a layered crystal structure consisting with the alternate ($Bi_2O_2$) layers and Se layers. The Bi-Se distance in $Bi_2O_2Se$ is 3.28 Å longer than the strong covalent bond length of 2.84-3.05 Å in $Bi_2Se_3$. It suggests the ($Bi_2O_2$) and Se layers are combined with relatively weak electrostatic interaction. Five possible defects are considered here, Se vacancies ($V_{Se}$), Bi vacancies ($V_{Bi}$), O vacancies ($V_O$), Se antisites at Bi positions ($Se_{Bi}$), and Bi antisites at Se positions ($Bi_{Se}$), as illustrated in Fig. 3a. The defect formation energies are calculated for the Se-rich and Se-poor conditions, as the displayed in Fig. 3b. The chemical potential of Se element at the Se-rich limit refers to the one in bulk Se, while the chemical potential of Se in $Bi_2Se_3$ is taken in consideration for the Se-poor limit (see Supplementary Information for details). Among five types of defects, $V_{Se}$ and $Se_{Bi}$ present the lowest formation energies at the Se-poor and Se-rich conditions, respectively. It indicates that $V_{Se}$ and $Se_{Bi}$ are probably the dominant defects in general. Because of the large slope of the $Se_{Bi}$ formation energy (Fig. 3b), the existence of $Se_{Bi}$ is expected to be strongly suppressed in the Se-poor condition. In short, $V_{Se}$ is the major defect in the Se-poor condition while $Se_{Bi}$ is the major one in the Se-rich condition.

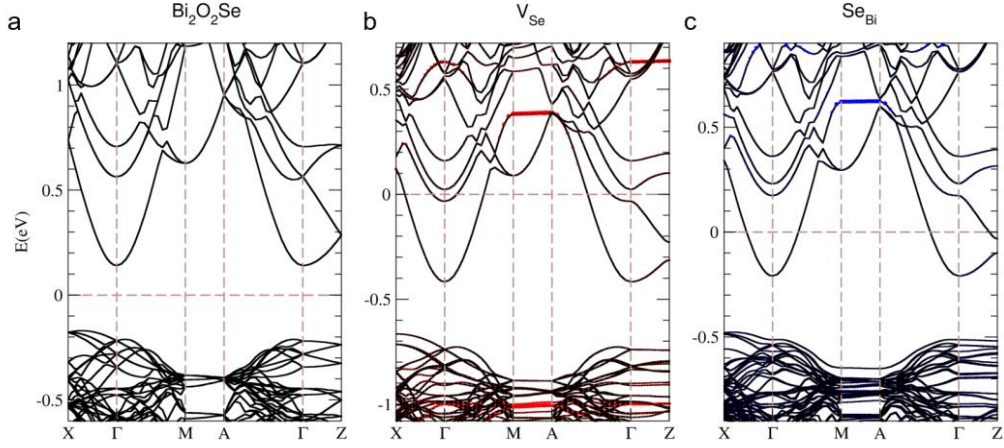

Figure 4. Calculated band structures of (a) perfect $Bi_2O_2Se$ and the configurations with domain defects (b) $V_{Se}$, (c) $Se_{Bi}$ antisite. Fermi level is set at 0. In (b), the red and blue dots respectively present the projected bands on the nearest neighbour Bi and Se atoms close to the $V_{Se}$. In (c), the blue dots denote the states located on the Se atom which is on a site before occupied by a Bi atom.

Next, we investigate the roles of $V_{Se}$ and $Se_{Bi}$ in doping the semiconductor $Bi_2O_2Se$. Figure 4 shows band structures of the pristine bulk and those with $V_{Se}$ and $Se_{Bi}$ defects. We choose a $5 \times 5 \times 2$ bulk supercell that contains 500 atoms to simulate the doping effect of a single defect, to suppress the interaction of defects from neighbored supercells. The pristine bulk presents an energy gap. The existence of both $V_{Se}$ and $Se_{Bi}$ shift the Fermi energy up into the conduction bands. In contrast to the doping effect of a conventional semiconductor (illustrated in Fig. 1a), both defects do not induce any in-gap donor states in the band structure. Corresponding donor states lie about 0.8 eV about the conduction band bottom. Each $V_{Se}$ ($Se_{Bi}$) site donates two (one) electrons to the bulk, in consistent with the shifted Fermi energy position. In addition, we note that the 0.32 eV indirect gap of $Bi_2O_2Se$ is underestimated by the density-functional theory (DFT). But the positions of $V_{Se}$ and $Se_{Bi}$ with respect to the conduction band are not affected by the DFT band gap. Similarly, $V_{Bi}$ and $V_O$ are found to be acceptors with acceptor levels in the valence bands (see Supplementary Information). The fact that $V_{Se}$ and $Se_{Bi}$ are dominant for different Se richness indicates corresponding samples are electron-doped, well consistent with our experiment observation.

Because the donor states caused by $V_{Se}$ and $Se_{Bi}$ are higher in energy than the conduction band bottom, excess electrons spontaneously move from donor states to the conduction band without requiring thermal activation. Therefore, the resultant electron carrier density does not increase exponentially with increasing temperature. i.e. $N_e \propto e^{-\Delta E/k_B T}$, where $\Delta E$ is the activation energy. Instead, the carrier density shows weak dependence on the temperature. It is consistent with the fact that the carrier density remains in the same order of magnitude from 2 K to 300 K, for both Se-rich and Se-poor samples (see Fig. 2b and 2d). Further, it also well explains that $R_{xx}$-$T$ behavior of $Bi_2O_2Se$ is similar to a metal rather than a semiconductor. As reducing the temperature, the decreasing resistance is mainly caused by the quickly increasing mobility.

Furthermore, we discuss different roles of $V_{Se}$ and $Se_{Bi}$ in the electron mobility. Recall that the wave function of the lowest conduction band distributes predominantly inside the ($Bi_2O_2$) layer, $V_{Se}$ defects that locate in the Se layer are spatially separated from the conducting electrons. Consequently, the scattering due to the ionized donor sites is suppressed, giving rise to the large mobility at low temperature. Here, the separation of donor sites and free electrons are naturally realized in a single phase material, leading to the same effect as the modulation doping. Thus, we refer to the high mobility mechanism in $Bi_2O_2Se$ as the self-modulation doping. For the other donor $Se_{Bi}$, the conducting channels are strongly disrupted because $Se_{Bi}$ sites locate inside the same ($Bi_2O_2$) layers, inducing much lower mobility than the $V_{Se}$ case. Therefore, the spatial distribution of two defects explains the large mobility variation between samples from the Se-poor ($V_{Se}$–dominated, high mobility) and Se-rich ($Se_{Bi}$–dominated, low mobility) conditions.

In conclusion, combing the experimental measurements and first-principles calculations, we have found that the self-modulation doping mechanism in the layered $Bi_2O_2Se$ semiconductor leads to the high electron mobility. Our work also explains the unusual metal-like resistance-temperature dependence of $Bi_2O_2Se$. Our findings paves a way to design high-mobility semiconductors in emerging layered materials.

**Acknowledgements**

B. H. Yan is supported by a research grant from the Benoziyo Endowment Fund for the Advancement of Science. H.L. Peng and J. X. Wu acknowledge financial support from the National Basic Research Program of China (Nos. 2014CB932500, 2016YFA0200101 and

2014CB920900), the National Natural Science Foundation of China (Nos. 21733001, 21525310 and 11774010), and China Postdoctoral Science Foundation Funded Project.

**Reference**:
[1] Dingle R, Störmer H L, Gossard A C, et al. Electron mobilities in modulation-doped semiconductor heterojunction superlattices. Applied Physics Letters, 1978, 33(7): 665-667.
[2] Yan B, Zhou G, Zeng X C, et al. Quantum confinement of crystalline silicon nanotubes with nonuniform wall thickness: Implication to modulation doping. Applied Physics Letters, 2007, 91(10): 103107.
[3] Yan B, Frauenheim T, Gali Á. Gate-controlled donor activation in silicon nanowires. Nano letters, 2010, 10(9): 3791-3795.
[4] Novoselov K S, Geim A K, Morozov S V, et al. Two-dimensional gas of massless Dirac fermions in graphene. Nature, 2005, 438(7065): 197.
[5] Zhang W, Huang Z, Zhang W, et al. Two-dimensional semiconductors with possible high room temperature mobility. Nano Research, 2014, 7(12): 1731-1737.
[6] Das S, Chen H Y, Penumatcha A V, et al. High performance multilayer $MoS_2$ transistors with scandium contacts. Nano letters, 2012, 13(1): 100-105.
[7] Liu H, Neal A T, Zhu Z, et al. Phosphorene: an unexplored 2D semiconductor with a high hole mobility. ACS Nano, 2014, 8(4): 4033-4041.
[8] Qiao J, Kong X, Hu Z X, et al. High-mobility transport anisotropy and linear dichroism in few-layer black phosphorus. Nature communications, 2014, 5: 4475.
[9] Kim J, Baik S S, Ryu S H, et al. Observation of tunable band gap and anisotropic Dirac semimetal state in black phosphorus. Science, 2015, 349(6249): 723-726.
[10] Liang T, Gibson Q, Ali M N, et al. Ultrahigh mobility and giant magnetoresistance in the Dirac semimetal $Cd_3As_2$. Nature materials, 2015, 14(3): 280.
[11] Shekhar C, Nayak A K, Sun Y, et al. Extremely large magnetoresistance and ultrahigh mobility in the topological Weyl semimetal candidate NbP. Nature Physics, 2015, 11(8): 645.
[12] Wu J, Yuan H, Meng M, et al. High electron mobility and quantum oscillations in non-encapsulated ultrathin semiconducting $Bi_2O_2Se$. Nature nanotechnology, 2017, 12(6): 530.
[13] Yin J, Tan Z, Hong H, et al. Ultrafast, highly-sensitive infrared photodetectors based on two-dimensional oxyselenide crystals. arXiv:1712.05942, 2017.
[14] Meng M, Huang S, Tan C, et al. Strong spin-orbit interaction and magnetotransport in semiconductor $Bi_2O_2Se$ nanoplates. Nanoscale, 2018, 10, 2704.
[15] Wu M, Zeng X C. Bismuth Oxychalcogenides: A New Class of Ferroelectric/Ferroelastic Materials with Ultra High Mobility. Nano letters, 2017, 17(10): 6309-6314.
[16] Hasan M Z, Kane C L. Colloquium: topological insulators. Reviews of Modern Physics, 2010, 82(4): 3045.
[17] Qi X L, Zhang S C. Topological insulators and superconductors. Reviews of Modern Physics, 2011, 83(4): 1057.

[18]   Taskin A A, Sasaki S, Segawa K, et al. Manifestation of topological protection in transport properties of epitaxial $Bi_2Se_3$ thin films. Physical review letters, 2012, 109(6): 066803.

[19]   Kim D, Cho S, Butch N P, et al. Surface conduction of topological Dirac electrons in bulk insulating $Bi_2Se_3$. Nature Physics, 2012, 8(6): 459.

[20]   Taskin A A, Sasaki S, Segawa K, et al. Achieving surface quantum oscillations in topological insulator thin films of $Bi_2Se_3$. Advanced Materials, 2012, 24(41): 5581-5585.

[21]   Thurmond C D. The standard thermodynamic functions for the formation of electrons and holes in Ge, Si, GaAs, and GaP. Journal of the Electrochemical Society, 1975, 122(8): 1133-1141.

[22]   Wu J, Tan C, Tan Z, et al. Controlled synthesis of high-mobility atomically thin bismuth oxyselenide crystals. Nano letters, 2017, 17(5): 3021-3026.

[23]   Radisavljevic B, Kis A. Mobility engineering and a metal–insulator transition in monolayer $MoS_2$. Nature materials, 2013, 12(9): 815.

[24]   Li L, Ye G J, Tran V, et al. Quantum oscillations in a two-dimensional electron gas in black phosphorus thin films. Nature nanotechnology, 2015, 10, 608.

[25]   Cui X, Lee G H, Kim Y D, et al. Multi-terminal transport measurements of $MoS_2$ using a van der Waals heterostructure device platform. Nature nanotechnology, 2015, 10, 534.

[26]   Kresse G, Furthmüller J. Efficient iterative schemes for ab initio total-energy calculations using a plane-wave basis set. Physical review B, 1996, 54, 11169.

**Supplementary** includes DFT method, experimental method, formation energy calculation for five defects, band structures for $Bi_2O_2Se$ with the defects $V_{Bi}$, $V_O$ and $Bi_{Se}$, and electronic properties for 3×3×1 supercell with spin-orbit coupling considered.

# Self-Modulation Doping Effect in the High-Mobility Layered Semiconductor Bi$_2$O$_2$Se

Huixia Fu[1] [*], Jinxiong Wu[2] [*], Hailin Peng[2], and Binghai Yan[1]

1. Department of Condensed Matter Physics, Weizmann Institute of Science, Rehovot, 7610001, Israel
2. Center for Nanochemistry, Beijing Sciences and Engineering Centre for Nanocarbons, College of Chemistry and Molecular Engineering, Peking University, Beijing 100871, China

Email: binghai.yan@weizmann.ac.il
* These two authors contributed equally to the work.

1. **DFT method** Density functional theory (DFT) calculations were performed using the Vienna ab initio simulation package (VASP) with core electrons represented by the projector-augmented-wave (PAW) potential. Plane waves with a kinetic energy cutoff of 300 eV were used as the basis set. A ($5 \times 5 \times 2$) supercell is chosen to simulate the vacancy in Bi2O2Se, which contains 200 Bi, 100 Se, 200 O atoms with lattice constant of 19.34 Å ×19.34 Å ×24.32 Å. The k-point grid of $1 \times 1 \times 1$ was used for Brillouin zone sampling. Geometry optimization was carried out until the residual force on each atom was less than 0.04 eV/Å. The qualitative results are also verified in the smaller calculation supercell of ($3 \times 3 \times 1$) with taking spin-orbit coupling into account. The details on the formation energy calculations are present as part 3 in the supplementary.

2. **Experimental method**

    a. **CVD growth of 2D Bi$_2$O$_2$Se crystals**

The 2D Bi$_2$O$_2$Se samples were obtained by a previously reported Bi$_2$Se$_3$-Bi$_2$O$_3$ co-evaporation method in a home-made low-pressure CVD system. The Bi$_2$O$_3$ powder (Alfa Aesar, 5N) and Bi$_2$Se$_3$ bulk (Alfa Aesar, 5N) were placed in the hot center and upstream 5 cm away as the source materials for evaporation, respectively. In our Se-richness controlled CVD growth experiments, the Se partial pressure is controlled by adjusting the ratio of Bi$_2$Se$_3$ and Bi$_2$O$_3$. In detail, we keep the amounts of Bi$_2$O$_2$Se constant as 2.0 g, while only changing the amount of Bi$_2$Se$_3$ from 2.0 g (relatively Se-rich) to 0.5 g (relatively Se-poor). The CVD growth process is fixed as follows: 1) raise the temperature to the targeted temperature of 650 ºC within 20 min, 2) keep the temperature of 650 ºC for 30 min, 3) cool down naturally for another 30 min and rapidly cool down to room temperature by electric fans. The growth substrates of freshly cleaved flurophlogopite mica were located at 10 cm away from the hot center. The carrier gas is Ar with a flow rate of 200 s.c.c.m. The pressure was kept constant at 400 Torr during CVD growth using a butterfly value.

### b. Device fabrication and electrical measurements

The non-encapsulating six-terminal Hall-bar devices of 2D Bi$_2$O$_2$Se were fabricated directly on the insulating mica substrates. First, location definition markers were predefined onto the mica using standard photolithography techniques. Then, electron-beam lithography (EBL) was used to write multiple metal contacts for both the six-terminal Hall-bar structures, followed by thermal evaporation of the contact metals Pd/Au (6/50 nm). The extra conductive protective layer (SX AR-PC-5000) was spin-coated on top of a PMMA layer to eliminate charge accumulation on the insulating mica substrate during EBL. The room temperature Hall measurements were performed in the four-probe configuration in the Physical Properties Measurement Systems (PPMS-9 T, Quantum Design), with magnetic field perpendicular to the devices.

## 3. Formation energy of five defects in Bi$_2$O$_2$Se

### a. When Se is rich:

In the experiment, bulk Se is used as the Se source instead of Bi$_2$Se$_3$. Thus we take the $\mu(\text{Se}) = \mu(\text{Se}_{\text{bulk}})$ as the chemical potential of element Se in our calculations.

The chemical reaction: Bi$_2$O$_3$+ 2Se → Bi$_2$O$_2$Se+ SeO$_2$ (gas) should meet the conditions as blow:

$$2\mu(\text{Bi}) + 2\mu(\text{O}) + \mu(\text{Se}) \geq E(\text{Bi2O2Se})$$
$$\mu(\text{Se}) + 2\mu(\text{O}) \geq E(\text{SeO2}_{\text{gas}})$$
$$2\mu(\text{Bi}) + 3\mu(\text{O}) \geq E(\text{Bi2O3})$$
$$2\mu(\text{Bi}) + 3\mu(\text{Se}) < E(\text{Bi2Se3})$$
$$\mu(\text{Bi}) < \mu(\text{Bi}_{\text{bulk}})$$
$$\mu(\text{O}) < \mu(\text{O}_{\text{O2 molecular}})$$

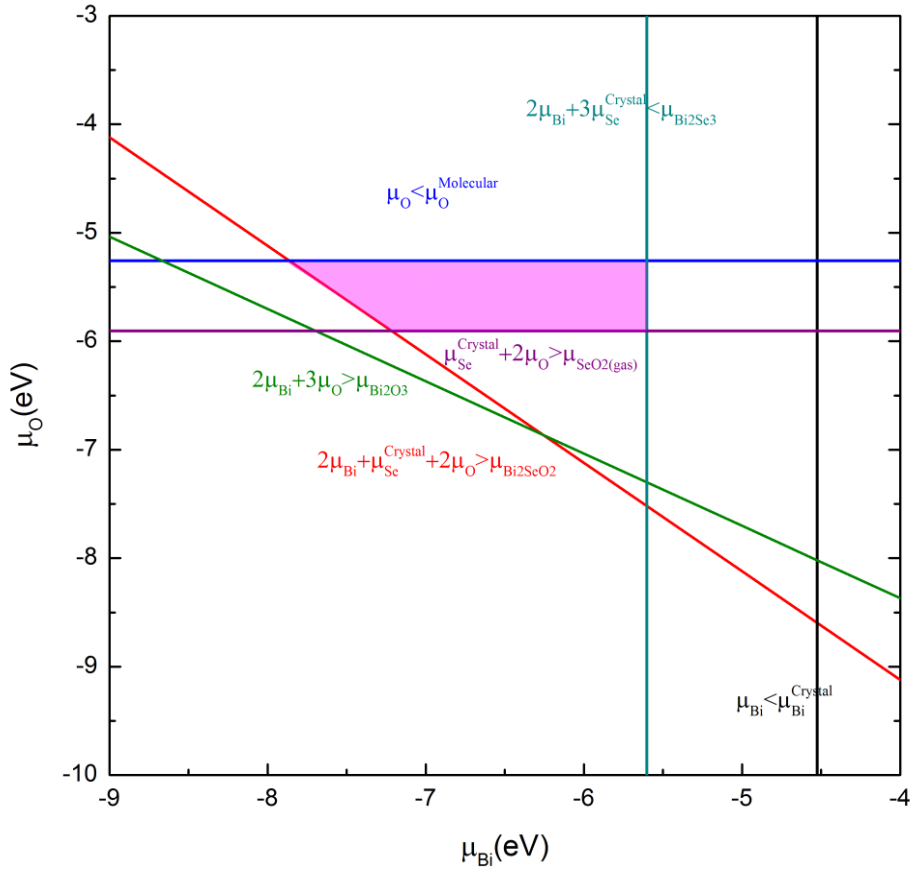

Figure S1. Atomic chemical potentials for $Bi_2O_2Se$ under Se-rich condition.

According to the thermodynamic equilibrium conditions, the chemical potentials of element Bi and O are limited in the desirable range as the pink area shown. The band gap in calculations is underestimated, thus we corrected the band gap as the experimental value. Band gap correction is defined as $\Delta Gap = 0.8 - 0.31 = 0.49$ eV. For electron donors, q is the number of electrons which locate at the energy range from CBM to Fermi level.

i) $\Delta H(VSe) = E(VSe\_supercell) + \mu(Se) - E(Bi2O2Se\_supercell) + q * \Delta Gap$
$V_{Se}$ is an electron donor (*n*-type) with q=2.

ii) $\Delta H(VBi) = E(VBi\_supercell) + \mu(Bi) - E(Bi2O2Se\_supercell)$
$V_{Bi}$ is an electron acceptor (*p*-type).

iii) $\Delta H(VO) = E(VO\_supercell) + \mu(O) - E(Bi2O2Se\_supercell) + q * \Delta Gap$
$V_O$ is an electron donor (*n*-type) with q=2.

$iv) \Delta H(SeonBi) = E(SeonBi\_supercell) + \mu(Bi) - \mu(Se) - E(Bi2O2Se\_supercell) + q * \Delta Gap$

$Se_{Bi}$ is an electron donor (*n*-type) with q=1.

$v) \Delta H(BionSe) = E(BionSe\_supercell) + \mu(Se) - \mu(Bi) - E(Bi2O2Se\_supercell)$

$Bi_{Se}$ is an electron acceptor (*p*-type).

b. When Se is poor:

Bi$_2$Se$_3$ is used as the Se source. Thus we take the limit $\mu(Se) = \frac{\mu(Bi2Se3) - 2 \times \mu(Bi_{bulk})}{3}$ as the chemical potential of element Se in our calculations.

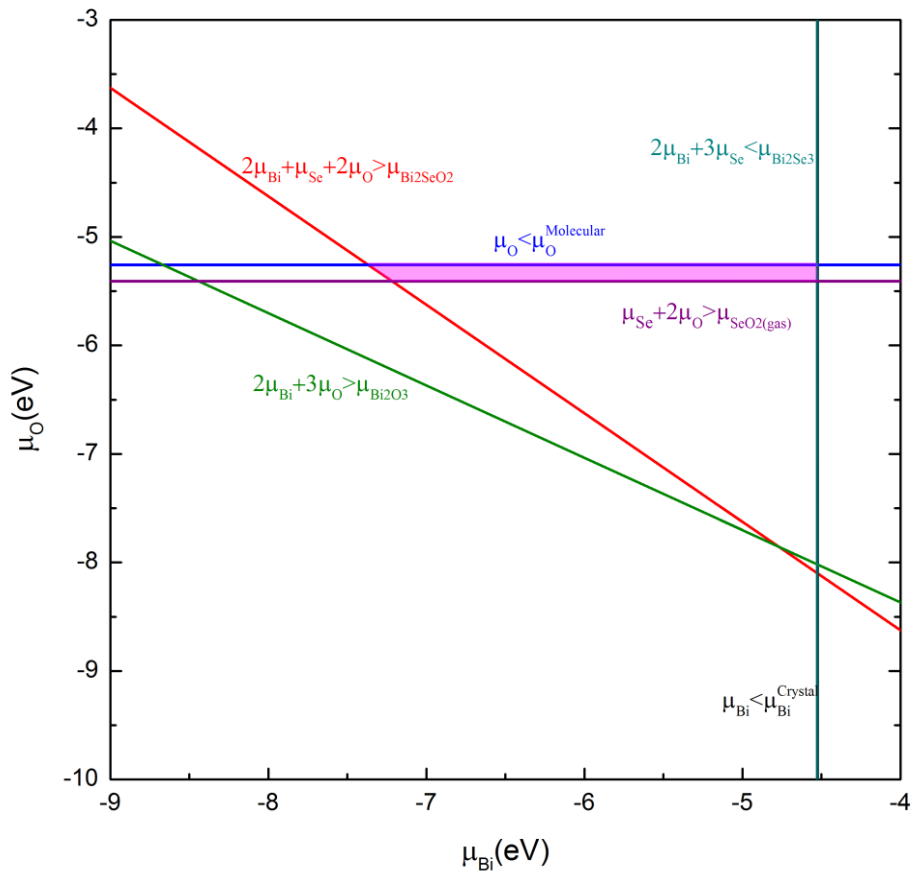

Figure S2. Atomic chemical potentials for Bi$_2$O$_2$Se under Se-poor condition. According to the thermodynamic equilibrium conditions, the chemical potentials of element Bi and O are limited in the desirable range as the pink area shown.

## 4. Band structures for Bi$_2$O$_2$Se with the defects V$_{Bi}$, V$_O$ and Bi$_{Se}$

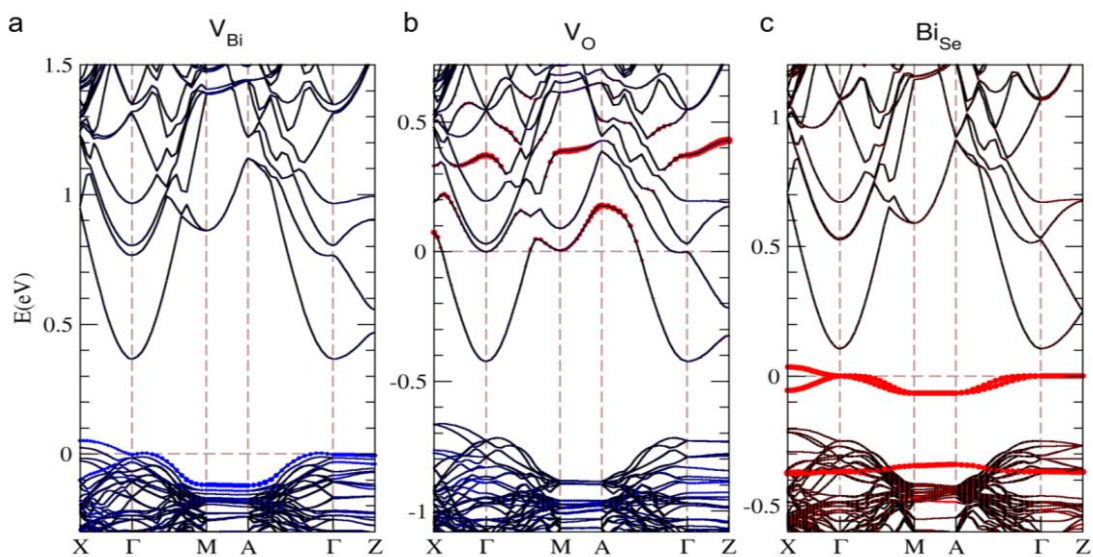

Figure S3. Calculated band structures of the Bi$_2$O$_2$Se configurations with the defects (a) V$_{Bi}$, (b) V$_O$ and (c) Bi$_{Se}$ antisite. Fermi level is set at 0. In (a) and (b), the red and blue dots respectively present the projected bands on the nearest neighbour Bi and Se atoms close to the defects. In (c), the red dots denote the states located on the Bi atom which is on a site before occupied by a Se atom.

## 5. Electronic properties for 3x3x1 supercell with spin-orbit coupling considered

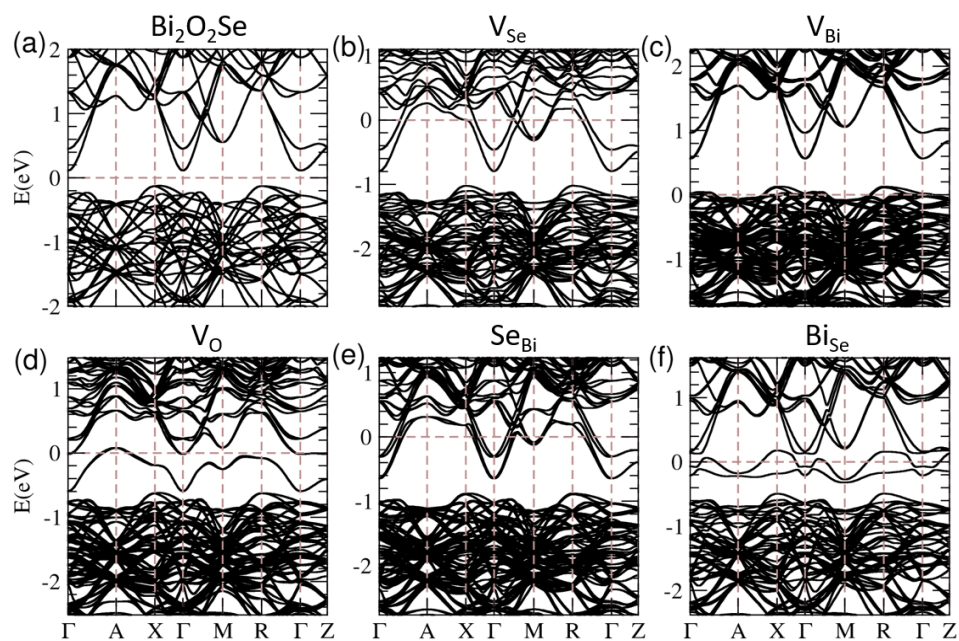

Figure S4. Calculated band structures of perfect $Bi_2O_2Se$ and the configurations with five defects. The $3 \times 3 \times 1$ supercell is chosen with spin-orbit coupling considered. Fermi level is set at 0.